# The 12 prophets dataset

(Aleijadinho 3D: technology for the preservation and dissemination of cultural heritage)

J Rodrigues, M Gazziro, N Gonçalves, O Neto, Y Fernandes, A Gimenes, C Alegre, R Assis
**University of Sao Paulo, Imprimate, and Leica Geosystems**

**http://aleijadinho3d.icmc.usp.br/data.html**

**Abstract**

The "Ajeijadinho 3D" project is an initiative supported by the University of São Paulo (Museum of Science and Dean of Culture and Extension), which involves the 3D digitization of art works of Brazilian sculptor Antonio Francisco Lisboa, better known as Aleijadinho. The project made use of advanced acquisition and processing of 3D meshes for preservation and dissemination of the cultural heritage. The dissemination occurs through a Web portal, so that the population has the opportunity to meet the art works in detail using 3D visualization and interaction. The portal address is http://www.aleijadinho3d.icmc.usp.br. The 3D acquisitions were conducted over a week at the end of July 2013 in the cities of Ouro Preto, MG, Brazil and Congonhas do Campo, MG, Brazil. The scanning was done with a special equipment supplied by company Leica Geosystems, which allowed the work to take place at distances between 10 and 30 meters, defining a non-invasive procedure, simplified logistics, and without the need for preparation or isolation of the sites. In Ouro Preto, we digitized the churches of Francisco of Assis, Our Lady of Carmo, and Our Lady of Mercy; in Congonhas do Campo we scanned the entire Sanctuary of Bom Jesus de Matosinhos and his 12 prophets. Once scanned, the art works went through a long process of preparation, which required careful handling of meshes done by experts from the University of São Paulo in partnership with company Imprimate.

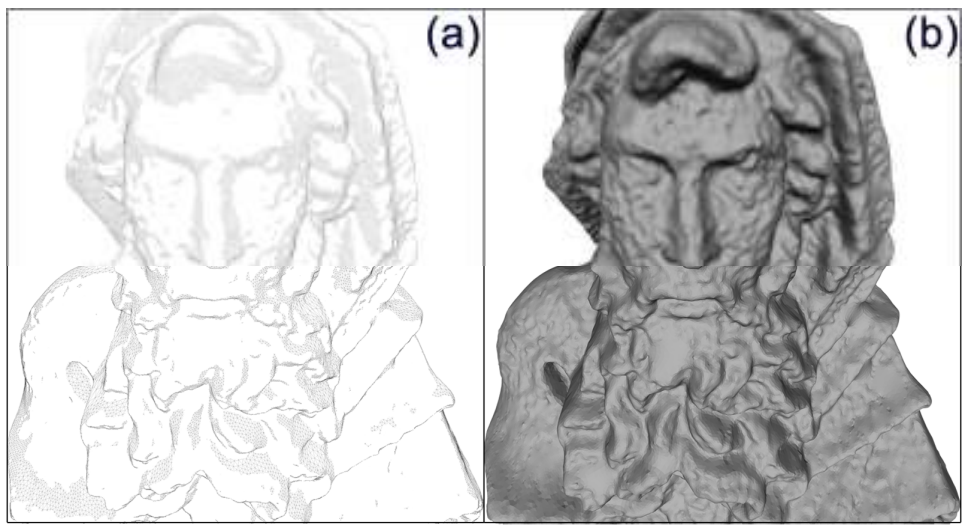

**FIGURE 1 – (a)** 3D mesh of prophet Isaias. **(b)** Mesh of points after triangulation.

**Introduction**

The "Aleijadinho 3D" project, conducted by the University of São Paulo (USP) and other partners, aims to disseminate, through a Web portal, the works attributed to Antonio Francisco Lisboa (1730-1814), better known as Aleijadinho. Considered one of the greatest exponents of the Brazilian colonial architecture, his history is full of gaps and controversies. Next to the bicentenary of his death, the only consensus seems to be about the originality and importance of his works, mostly carved in wood and soapstone. For this project, some of his works located outdoors were chosen for scanning; these works are more prone to suffer a deterioration process, what increases the importance of the preservation. Specifically, we selected the Churchyard of the Sanctuary of Bom Jesus de Matosinhos in Congonhas do Campo, MG, Brazil; and the churches in Ouro Preto, MG, Brazil: churches of Francisco of Assis, Our Lady of Carmo, and Our Lady of Mercy.

**3D Digitizing**

The use of 3D scanning technology has been more frequently used in the last decade, however, it is still a new technology that requires experts and special procedures for their use. Some examples of 3D digitization are the Michelangelo's David (Levoy, 2000), and the scan of the very Sanctuary of Bom Jesus de Matosinhos, conducted by the Federal University of Paraná (UNESCO) that, in 2011, scanned the prophets site for preservation purposes.

The process of 3D scanning is quite complex compared to the acquisition of 2D pictures. It requires that the works be scanned from different angles (front, sides, back, above and below), a process that takes time and precision electronics and mechanics. The scanning of a single statue, for example, may take from a few hours to a full day, depending on the technology that is used. In computational terms, each scanning produces one file; hence, a single statue can produce more than a dozen files, each containing one fraction of the 3D information. These files then need to be combined to compose a single three-dimensional information.

The three-dimensional information of each piece is produced initially as a set of points in space, which is called three-dimensional mesh. This mesh is the basis of all the work to be done afterwards, whose goal is to computationally reproduce a work of the real world. Once the mesh is available in a single combined file, the next step is called triangulation. Roughly speaking, triangulation is the process of surrounding a mesh of points with a surface; it is this process that makes the 3D scanning visualizable, because it turns the mesh into an object that reflects light and that can be perceived in the same way we perceive things in the physical world.

Triangulation is a process which, as its very name implies, is based on triangles - see Figure 1, so that the higher the number of triangles, the better the quality of the object being visualized. However, large numbers of triangles are prohibitive for visualization, because the computational processing is as great as the number of triangles. Indeed, in this project, the initial number of triangles was large enough to prevent the proper management of the data; even with powerful computers and large amounts of memory, the collected data made the process too slow. This restriction is also decisive in respect to the dissemination of works digitized in 3D - a huge number of triangles cannot be efficiently accessed via the Web, and even if it were, special computers would be needed for the data to be visualized. These limitations led to the need for a data preparation step named sampling (or decimation). Sampling analyzes the original data, discarding information that is constituent of the art work, but that can be discarded without prejudice. As an illustration, we refer to a similar process carried out with digital music; when it is transformed into the popular MP3 format, it goes through a process of disposal of sounds without prejudice to the quality of the music. The same process is applied when working with digitized works in 3D and it was done in this project.

In what concerns the digitization of art works exhibited outdoors and in large scales (such as churches and statues), the process presents another challenge. When the scan is performed, the generated data carry problems arising from the format and location of the art works, such as the statues. Some parts of the art pieces are not adequately exposed and accessible to the scanning equipment; besides that, the objects and their parts are positioned one in front of each other, generating occlusion, regions of "shadow" that the scanning equipment cannot reach. As result, the mesh of points, after combination, triangulation, and sampling, presents "holes" - literally; pieces of the artwork that are visualized as black holes that ruin the appearance of objects. Thus, unlike what you have with 2D photographs, the works are not ready after being exposed to the equipment. In fact, most of the work occurs afterwards, when the holes are patched, and missing parts of the work must be manually remodeled.

Another issue is the fact that the 3D meshes, as they were digitized in this project, are not colored. Nearby points of the meshes are only spatial information without color that, when triangulated, create objects with metallic appearance not faithful to the original art work. Thus, the next step - after the combination, triangulation, sampling, and fixing of the mesh - is the coloring process. Once the 3D works are consistent and without holes, it is necessary to assign to each of its parts the look, and not just the shape, of the original work. For this step, we needed pictures from different angles of each work (back and front), which are used to define the appearance of each triangle in the 3D visualization - see Figure 2. Note that there exists a similar process, named texturing; differently, in this project, we used a process named coloring, which brings more realistic results.

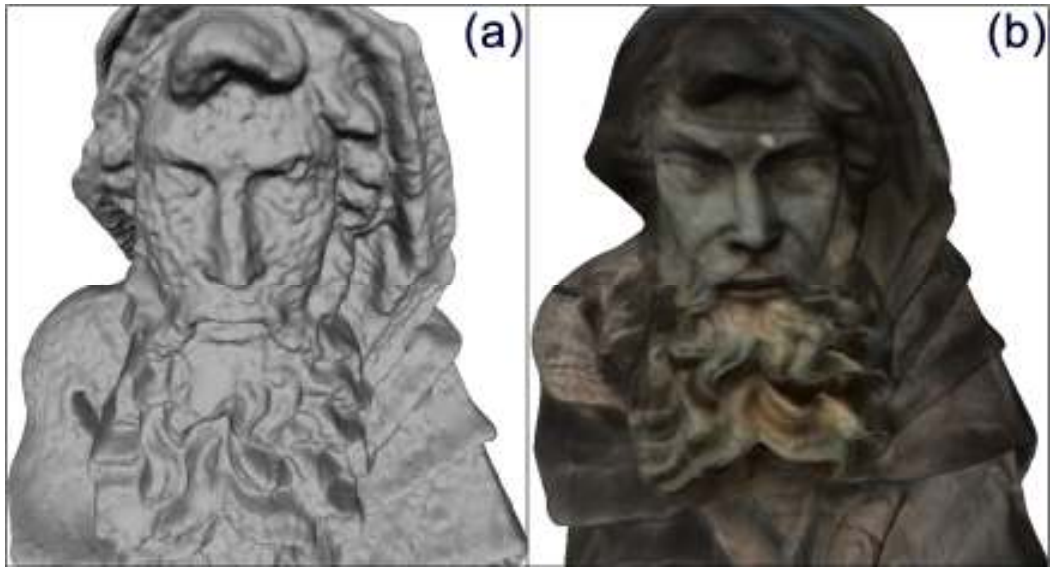

**FIGURE 2 – (a) Statue of the prophet Isaias before coloring. (b) Colored result.**

In this project, each part of the art work - statues, churches, environment, and adornment - was treated separately until the coloring process; this was done for the purpose of simplifying the process and also to have different parts processed in parallel by different professionals. After coloring, the 3D art works are visualizable in a format very close to the original pieces; now it is necessary to unite all the pieces in a single environment. This is the last step, which prepares the final outcome; this step leads to adjustments in all previous steps, because when the data are put together, one can perceive imperfections and further needs of sampling, correction, and coloring. Also at this stage, the 3D works are combined in a software capable of promoting 3D interaction; in the case of this project, the works were placed in an environment where you can walk, jump, and look in all directions, a feature that requires specialized technology.

**Execution**

Initially, the project called for the use of low-cost handheld 3D scanners, which digitize the works at distances between 1 and 3 meters. This characteristic would require the closing of the sites where the art works lie so to have controlled conditions. However, due to bureaucratic matters, we were not allowed to work this way. The alternative was the use of a high-cost special equipment capable of scanning the artworks remotely (between 10 and 30 meters) and that worked even with the traffic of tourists and curious – refer to Figure 3. This equipment was supplied by company Leica Geosystems (leica-geosystems.com.br), collaborator on this project.

With the proper equipment, the art works of Congonhas were scanned in one day, including not only the 12 prophets, but also the facade of the church of the sanctuary and its adornment. Despite the suitability of the equipment, the process was quite difficult due to the constant presence of people between the equipment and the works; consequently, later we had a lot more of work on remodeling the missing parts of the scanned objects, what took nearly 3 months to complete. In addition, some details of the works were not collected. The same work was performed in the city of Ouro Preto, in the churches of Francisco of Assis, Our Lady of Carmo, and Our Lady of Mercy, in a total of 4 days.

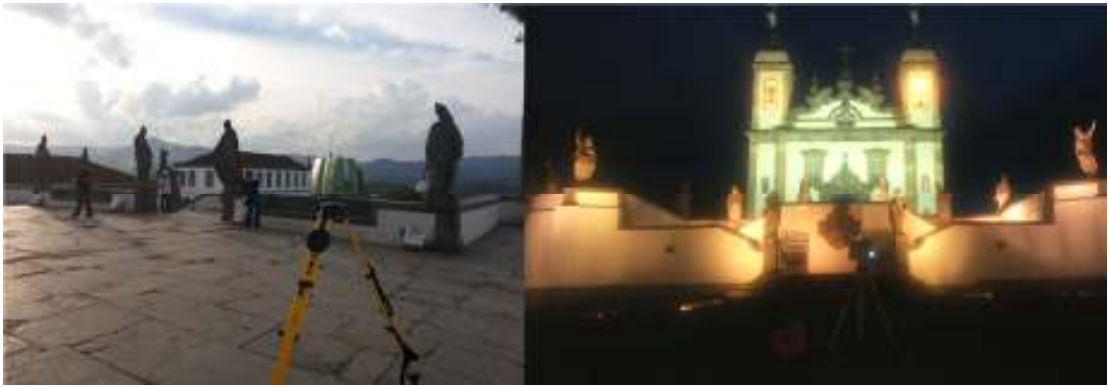

**FIGURE 3 – The digitizing process in the city of Congonhas do Campo, MG, Brazil.**

With the 3D raw meshes, we began the process of preparing the data. We used open source software in the steps of combination, triangulation, sampling and correction of the meshes. Software MeshLab (http://meshlab.sourceforge.net/) was used for data manipulation: viewing, cutting, hole filling, partitioning, and smoothing of meshes, among other tasks. Software Blender (http://www.blender.org/) was used for modeling, sampling, repositioning, and combination of the art works, among other tasks. The preparation of the 3D meshes, as already mentioned, was the most costly step of the process, not only because it is a laborious activity, but also because it is highly complex. It demanded the project team to learn new skills and techniques, passing through many stages of trial and error. The work was much higher than initially expected.

The next step was the coloring of the 3D works. This step initially demanded pictures of each of the original art works, front and back. At this point, company Imprimate (http://www.imprimate.com.br/) joined the project as a second collaborator. This partnership was needed because the coloring process of 3D meshes from photographs is very complex, requiring not only extensive experience as well as special software that is extremely expensive. Thus, the coloring process was used in place of the, simpler, texturing process originally envisaged for the project. The results were more realistic.

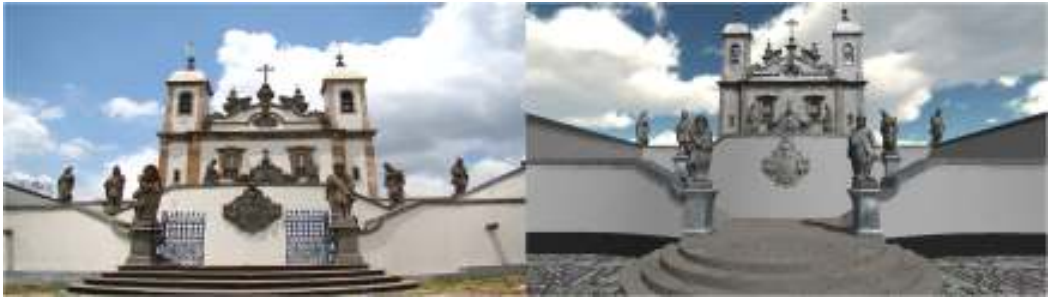

**FIGURE 4 – Website of the Project, with a picture of the original site in Congonhas do Campo; comparatively, at the right, one can see the synthetic 3D version developed in this project.**

The last step was the union of the different parts of the works, especially the art works of Congonhas - Figure 4. The site of the city of Congonhas was divided into 17 parts: 12 Prophets, Adrio, coat, door, door adornment, and church facade. These parts were processed in parallel and assembled to form the original environment. The combination of the parts was performed using software Unity (http://unity3d.com/), which allows the edition of 3D projects, and also an interaction mechanism with many functionalities. In this project, Unity is used so that the user can get around the 3D environment in which he/she can look in all directions, virtually exploring the work. Figure 5 illustrates this possibility in an environment of a digital cave, although this type of environment is not required; actually, any user can interact with the system in a conventional personal computer.

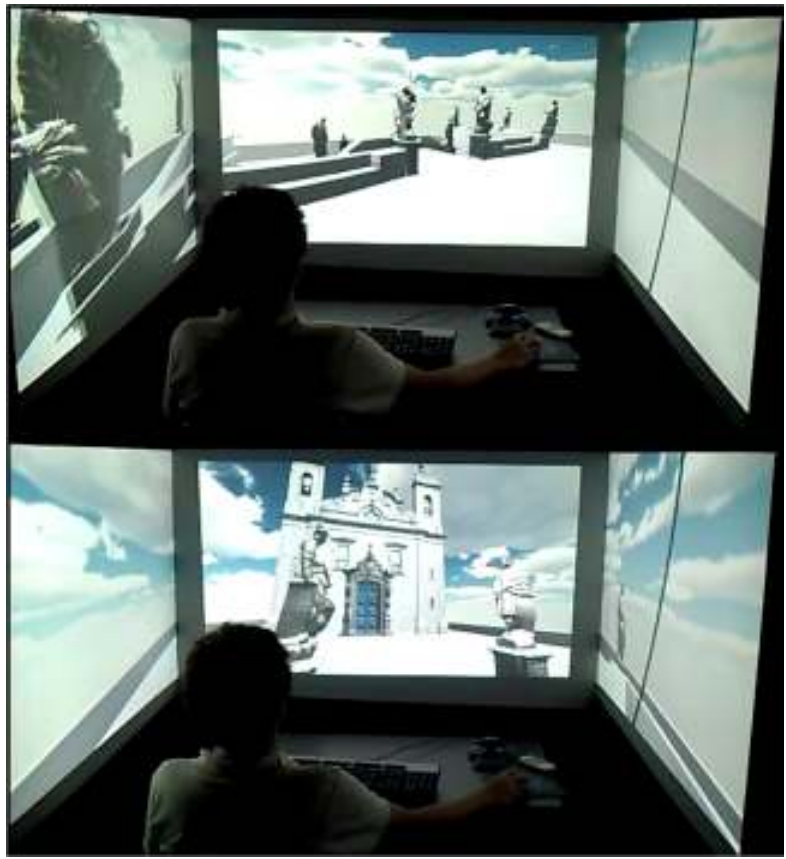

**FIGURE 5 – Interactive result of the project visualized in a digital cave.**

### Dissemination via Web site

As part of the project, we built a website accessible at http://aleijadinho3d.icmc.usp.br, illustrated in Figure 6. The site contains information about Aleijadinho's work provided along with photos, videos, information on the project, and with the results of the 3D scanning. The goal of the site is to spread the innovation of the work, demonstrating how to reproduce a cultural heritage using computer technology. The dissemination is a cultural initiative of the University of São Paulo, supported by their institutions Museum of Science, and Dean of Culture and Extension.

### Datasets

The datasets produced in this work are available for research purposes on the Web site. The data are distributed in Polygon File Format (ply) after all the processing described in the 3D Digitizing section. We encourage researchers and professionals to download, experiment, and investigate our data.

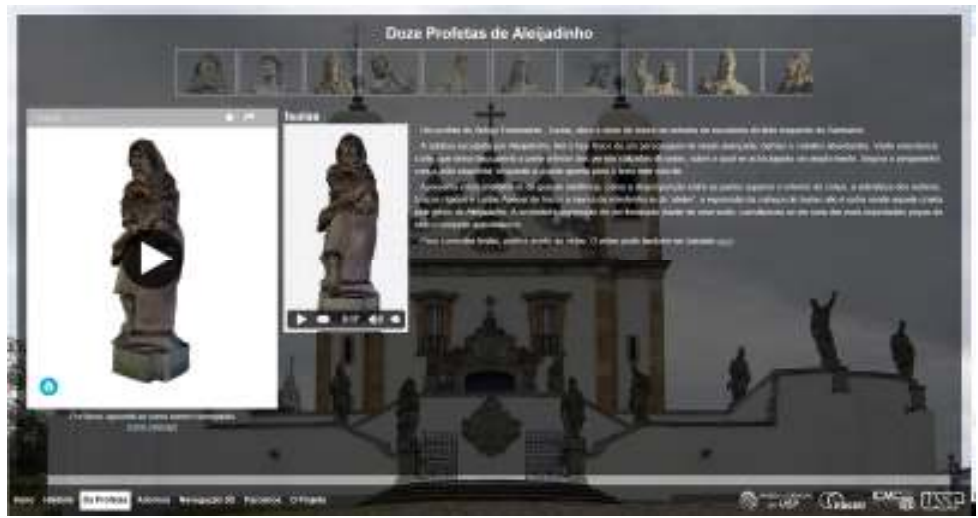

**FIGURE 6 – Website of the Project: http://aleijadinho3d.icmc.usp.br.**

**Conclusions**

This paper described the project of 3D digitization and dissemination of the art works of the Brazilian artist Aleijadinho in the cities of Congonhas do Campo, MG, Brazil, and Ouro Preto, MG, Brazil. We described the technical processes and also the stages of the project implementation, including the difficulties and solutions. The result was the creation of a computational version of the artist's work using 3D technology; the works were made available in a Web site (http://aleijadinho3d.icmc.usp.br), where users can virtually visit the art works and interact with immersive computing environments, simulating the visitation to the original sites. It is expected that the results will lead to the widespread dissemination of Aleijadinho's work, enriching the cultural experience of people from throughout Brazil and abroad. It is also expected that the results will serve as bases for the preservation of the art works.